\documentclass[%
 aip,
 prb,
 sd,%
 amsmath,amssymb,
 preprint]{revtex4-1}

\usepackage{graphicx}% Include figure files
\usepackage{dcolumn}% Align table columns on decimal point
\usepackage{bm}% bold math
\usepackage{color}% bold math

\begin{document}

\title{
Critical current density of a spin-torque oscillator with an in-plane
magnetized free layer  and an out-of-plane magnetized polarizer
}

\author{R. Matsumoto}
\author{H. Imamura}\email{h-imamura@aist.go.jp}
\affiliation{ 
$^{1}$National Institute of Advanced Industrial Science and Technology (AIST),
Spintronics Research Center, Tsukuba, Ibaraki 305-8568, Japan
}

\date{\today}

\begin{abstract}
Spin-torque induced magnetization dynamics in a spin-torque oscillator 
with an in-plane (IP) magnetized free layer and 
an out-of-plane (OP) magnetized polarizer 
under IP shape-anisotropy field ($H_{\rm k}$) and applied IP magnetic
field ($H_{\rm a}$) was theoretically studied based on the macrospin model. 
The rigorous
analytical expression of the critical current density
($J_{\rm c1}$) for the OP precession was obtained. The obtained expression 
successfully reproduces the experimentally obtained
$H_{\rm a}$-dependence of $J_{\rm c1}$ reported in [D. Houssameddine 
$et$ $al$., Nat. Mater. 6, 447 (2007)].
\end{abstract}


\maketitle

\begin{figure}[t]
%\begin{figure}[P]
\includegraphics{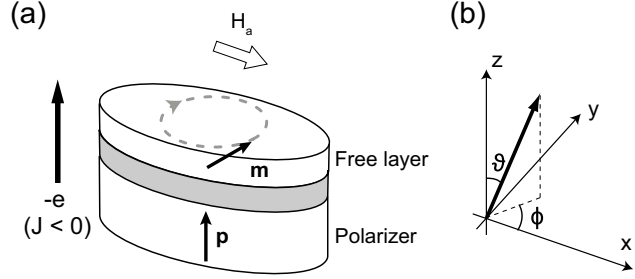}% Here is how to import EPS art
\caption{\label{fig:model} 
  (a)
  Spin-torque oscillator consisting of in-plane (IP) magnetized free layer
  and out-of-plane (OP) magnetized polarizer layer.
  IP magnetic field ($H_{\rm a}$) is applied parallel to easy axis of
  the free layer.   Negative current density $(J<0)$ is defined as
  electrons flowing from the polarizer layer to the free layer.
  The unit vector $\bm{m}$ represents the direction of magnetization
  in the free layer.
  (b) 
  Definitions of Cartesian coordinates $(x, y, z)$, 
  polar angle ($\theta$) and azimuthal angle ($\phi$).
\label{fig:fig1}
}
\end{figure}

%========================================
% Introduction
%========================================
A spin-torque oscillator (STO)
\cite{slonczewski_current-driven_1996,berger_emission_1996,tsoi_excitation_1998,kiselev_microwave_2003,deac_bias-driven_2008,slavin_nonlinear_2009} 
with an in-plane (IP) magnetized free layer and 
an out-of-plane (OP) magnetized polarizer 
\cite{lee_analytical_2005, houssameddine_spin-torque_2007, firastrau_state_2007, silva_theory_2010, ebels_macrospin_2008, suto_magnetization_2012, lacoste_out--plane_2013, bosu_reduction_2016, hiramatsu_magnetic_2016} 
has been attracting a great deal of attention 
as microwave field generators
\cite{zhu_bias-field-free_2006, zhu_microwave_2008, wang_media_2009, igarashi_effect_2009, suto_magnetization_2012, kudo_resonant_2015}
and high-speed field sensors
\cite{kudo_numerical_2010, braganca_nanoscale_2010, suto_real-time_2011, rippard_spin-transfer_2010, kubota_spin-torque_2013, suto_nanoscale_2014}.
The schematic of the STO is illustrated in Fig. \ref{fig:fig1}(a).
When the current density ($J$) of the applied dc current exceeds the
critical value ($J_{\rm c1}$), the 360$^{\circ}$ in-plane precession of
the free layer magnetization,
so-called OP precession, is induced by the spin torque. 
Thanks to the OP precession, a large-amplitude microwave
field can be generated,
\cite{suto_magnetization_2012, bosu_reduction_2016, hiramatsu_magnetic_2016}
and a high microwave power can be obtained through the additional
analyzer.\cite{houssameddine_spin-torque_2007}

The critical current density,
$J_{\rm c1}$, for the OP precession
of this type of STO has been extensively studied both experimentally
\cite{houssameddine_spin-torque_2007,hiramatsu_magnetic_2016} 
and theoretically.
\cite{lee_analytical_2005, firastrau_state_2007, silva_theory_2010, morise_stable_2005, ebels_macrospin_2008,lacoste_out--plane_2013} 
In 2007, D. Houssameddine $et$ $al$. experimentally found that 
$J_{\rm c1}$ was approximately expressed as
$J_{\rm c1} \propto H_{\rm k} + 2 H_{\rm a} $
where 
$H_{\rm k}$ is IP shape-anisotropy field and
$H_{\rm a}$ is the applied IP magnetic field.
In theoretical studies, 
the effect of $H_{\rm k}$ and $H_{\rm a}$ on $J_{\rm c1}$  has been studied analytically and numerically.
U. Ebels $et$ $al$. proposed an 
apporximate expression of $J_{\rm c 1}$, however, as we shall show later, 
it gives exact solution only 
in the limit of $H_{\rm a} = 0$ and $H_{\rm k} \rightarrow 0$.
Lacoste $et$ $al$. obtained the lower current boundary for the existence of
OP precession
\cite{lacoste_out--plane_2013} which gives some insights
into $J_{\rm c1}$, however, it could be lower than $J_{\rm c1}$. To our
best knowledge, $J_{\rm c1}$ of this type of STO is still controversial
and a systematic understanding of $J_{\rm c1}$ in the presence of $H_{\rm
k}$ and $H_{\rm a}$ is necessary.

%==============================
% Fig. 1
%==============================

%========================================
% In this letter
%========================================
In this letter, 
we theoretically analyzed
spin-torque induced magnetization dynamics in the STO 
with an IP magnetized free layer and an OP magnetized polarizer
in the presence of $H_{\rm k}$ and $H_{\rm a}$
based on the macrospin model. 
We obtained the rigorous analytical expression of $J_{\rm c1}$ and
showed that it successfully reproduces the experimentally obtained
$H_{\rm a}$-dependence of the critical current reported by
D. Houssameddine $et$ $al$.
\cite{houssameddine_spin-torque_2007}

%========================================
% Model
%========================================
The system we consider is schematically illustrated in
Figs. \ref{fig:fig1}(a) and (b).
The shape of the free layer is either a circular cylinder or an
elliptic cylinder.
The lateral size of the nano-pillar is assumed to be so small that the
magnetization
dynamics can be described by the macrospin model. 
Directions of the magnetization in the free layer and in the polarizer are
represented by the unit vectors
$\bm{m}$ and $\bm{p}$, respectively.
The vector $\bm{p}$ is fixed to the positive $z$-direction.
The negative current is defined as electrons flowing
from the polarizer to the free layer.
The applied IP magnetic field, $H_{\rm a}$, 
is assumed to be parallel to the magnetization easy axis of the free layer.
The easy axis is parallel to $x$-axis.

The energy density of the free layer is given by
\cite{stiles_spin-transfer_2006}
\begin{align} \label{eq:E}
E = 
 & 
  \frac{1}{2} \mu_{0} M_{\rm s}^{2} ( N_{x} m_{x}^{2} + N_{y} m_{y}^{2} + N_{z} m_{z}^{2} )  \nonumber\\  
 % &-  \frac{1}{2} \mu_{0} M_{\rm s} H_{\rm k} \sin^{2} \theta  \cos^{2} \phi 
 & + K_{\rm u1} \sin^{2} \theta - \mu_{0} M_{\rm s} H_{\rm a}  \sin \theta  \cos \phi. 
 %- h_{{\rm a}z} \cos \theta.
\end{align}
Here 
($m_{x}$, $m_{y}$, $m_{z}$) = ($\sin \theta \cos \phi$, $\sin \theta \sin \phi$, $\cos \theta$), 
and
$\theta$ and $\phi$ are the polar and azimuthal angles of $\bm{m}$ 
as shown in Fig. \ref{fig:fig1}(b). The demagnetization coefficients, 
$N_{x}$, $N_{y}$, and $N_{z}$ are assumed to satisfy $N_{z} \gg N_{y}
\geq N_{x}$.
$K_{\rm u1}$ is the first-order crystalline anisotropy constant, 
$\mu_{0}$ is the vacuum permeability, 
$M_{\rm s}$ is the saturation magnetization of the free layer, 
and $H_{\rm a}$ is applied IP magnetic field.

Hereafter we conduct the analysis with dimensionless expressions. 
The dimensionless energy density of the free layer is given by
\begin{align} \label{eq:epsilon}
\epsilon = &
 \frac{1}{2} ( N_{x} m_{x}^{2} + N_{y} m_{y}^{2} + N_{z} m_{z}^{2} )  \nonumber\\  
 & + k_{\rm u1}  \sin^{2} \theta  
 %-  \frac{1}{2} h_{\rm k} \sin^{2} \theta  \cos^{2} \phi 
 - h_{\rm a}  \sin \theta  \cos \phi.
\end{align}
Here, 
$k_{\rm u1}$ and $h_{\rm a}$ are defined as 
$k_{\rm u1}=K_{\rm u1}/(\mu_{0} M_{\rm s}^{2})$
and $h_{\rm a} = H_{\rm a}/M_{\rm s}$.
We discuss on the spin-torque induced magnetization dynamics
at $h_{\rm a} \geq 0$ in this letter, 
however,  the dynamics 
at $h_{\rm a} <  0$ can be calculated in the similar way.

%========================================
% LLG equation
%========================================
The spin-torque induced dynamics of $\bm{m}$ in the presence
of applied current is described by the following Landau-Lifshitz-Gilbert equation,
\cite{stiles_spin-transfer_2006}
\begin{align} 
\label{eq:LLGtheta}
(1+\alpha^{2})
\frac{d\theta}{d\tau} 
&=  h_{\phi} + \chi \sin \theta + \alpha  h_{\theta}, \\
\label{eq:LLGphi}
(1+\alpha^{2})
\sin \theta \frac{d\phi}{d\tau} 
& = - h_{\theta}  + \alpha (h_{\phi} +\chi \sin \theta), 
\end{align}
where $\tau$, $\chi$, $h_{\theta}$, and $h_{\phi}$ are the dimensionless quantities
representing time, spin torque, and $\theta$, $\phi$ components of effective magnetic field, ${\bm h}_{\rm eff}$,
respectively. 
${\bm h}_{\rm eff}$ is given by ${\bm h}_{\rm eff}=-\nabla \epsilon$.
$\alpha$ is the Gilbert damping constant.  
The dimensionless time is defined as $\tau=\gamma_{0}M_{\rm s} t$,
where 
$\gamma_{0}=2.21 \times 10^{5}$ m/(A$\cdot$s)
is the gyromagnetic ratio and $t$ is the time.
$h_{\theta}$ and $h_{\phi}$ are given by
\begin{align} 
\label{eq:htheta}
h_{\theta} 
 =& \cos \theta \left[ 2 \sin \theta  
\left(\frac{h_{\rm k}}{2} \cos^{2} \phi - k_{\rm u1}^{\rm eff}\right) +  h_{\rm a} \cos \phi \right], \\
\label{eq:hphi}
h_{\phi}
 =&  - \frac{h_{\rm k}}{2} \sin \theta \sin 2\phi -  h_{\rm a} \sin \phi.
\end{align}
Here 
$h_{\rm k}$
is dimensionless IP shape-anisotropy field
being expressed as
$h_{\rm k}  = N_{y} - N_{x} = H_{\rm k}/M_{\rm s} $.
$k_{\rm u1}^{\rm eff}$ is defined as 
$k_{\rm u1}^{\rm eff} = K_{\rm u1}^{\rm eff}/(\mu_{0} M_{\rm s}^{2}) = k_{\rm u1} - (N_{z} - N_{y})/2$.
$K_{\rm u1}^{\rm eff}$ is the effective first-order anisotropy constant
where the demagnetization energy is subtracted.
Since we are interested in the spin-torque induced magnetization dynamics 
of the IP magnetized free layer, we concentrate on $k_{\rm u1}^{\rm eff} < 0 $.
The prefactor of the spin-torque term, $\chi$, is expressed as
\begin{equation}
 \label{eq:chi}
 \chi = 
 \eta(\theta)
  \frac{\hbar}{2 e}
  \frac{ J }{ \mu_{0} M_{\rm s}^{2} d},
\end{equation} 
where $\eta(\theta) = P/(1+P^{2} \cos \theta)$ is spin-torque efficiency, $P$ is the spin polarization, $J$ is the applied current density, 
$d$ is the thickness of the free layer,
$e\, (>0)$ is the elementary charge and $\hbar$ is the Dirac constant. 
For convenience of discussion, the sign of Eq. \eqref{eq:chi} is taken to
be opposite to that in Ref. \citenum{stiles_spin-transfer_2006}.

%==============================
% Fig. 2
%==============================
\begin{figure}[t]
%\begin{figure}[P]
\includegraphics[width=0.95\columnwidth]{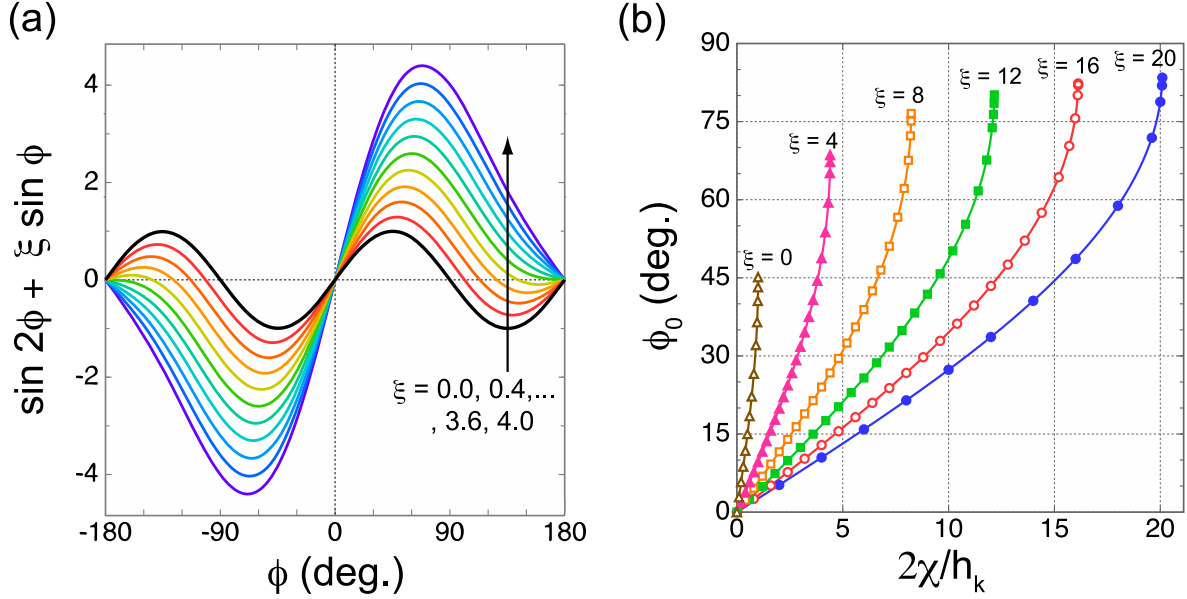}% Here is how to import EPS art
\caption{\label{fig:FixedPoint} 
  (a) 
  Function, $\sin 2\phi  + \xi \sin \phi$, is plotted as against $\phi$. 
  Value of $\xi$ is varied from 0.0 to 4.0.
  (b)
  Spin-torque magnitude ($\chi$) dependence of 
  $\phi$ at fixed point ($\phi_{0}$) 
  in the presence of IP shape anisotropy field. 
  $\chi$ is defined in Eq. (\ref{eq:chi}), and it is proportional to $J$. 
  Curves represent the analytical results obtained by Eq. \eqref{eq:Sin2phiSinphi}.
  Open or solid circles, squares, and triangles represent numerical calculation results. 
}
\end{figure}

In the absence of the current, i.e., $J=0$, the angles of the
equilibrium direction of $\bm{m}$ are obtained as
$\theta_{\rm eq}=\pi/2$ and $\phi_{\rm eq}=0$ by minimizing
$\epsilon$ with respect to $\theta$ and $\phi$. 
Application of $J$ changes $\theta$ and $\phi$ from its equilibrium
values. If the magnitude of $J$ is smaller than the critical value, 
the magnetization converges to a certain fixed point.
\cite{bertotti_preface_2009}
The equations 
determining the polar and azimuthal angles of the fixed point ($\theta_{0}$,
$\phi_{0}$) are obtained by setting $d\theta/d\tau=0$ and
$d\phi/d\tau=0$ as
\begin{align}
  h_{\theta}^{0} 
  &= 0,
  \\
  h_{\phi}^{0} 
  &= -\chi \sin \theta_{0}.
\end{align}
The fixed point around the equilibrium direction ($\theta_{\rm eq}=\pi/2$,
$\phi_{\rm eq}$=0) are obtained as follows. Assuming
$\left|\phi_{0}\right| \le \pi/2$, i.e., $\cos\phi_{0}\ge 0$ and noting
$k_{\rm u1}^{\rm eff}<0$, one can see that the quantity in the square
bracket of Eq. \eqref{eq:htheta} is positive and $\theta_{0}=\pi/2$ to
satisfy $h_{\theta}^{0} = 0$.
Substituting $\theta_{0}=\pi/2$ to $h_{\phi}^{0} = -\chi \sin \theta_{0}$,
the equation determining $\phi_{0}$ is obtained as
\begin{align}
\label{eq:Sin2phiSinphi}
 \sin 2\phi_{0}  + \xi \sin \phi_{0}  = 2 \chi/h_{\rm k},
\end{align}
where $\xi=2 h_{\rm a}/h_{\rm k}$.
Since Eq. \eqref{eq:Sin2phiSinphi} does not contain the Gilbert damping
constant, $\alpha$, $\phi_{0}$ is independent of $\alpha$.
In Fig. \ref{fig:FixedPoint}(a), the function, $\sin 2\phi  + \xi \sin \phi$,
is plotted against $\phi$ for various values of $0 \le \xi \le 4$.
One can clearly see that the azimuthal angle of the maximum (minimum)
increases (decreases) towards $\pi/2$ ($-\pi/2$) with increase of $\xi$.
The azimuthal angle of the fixed point is given by the intersection
of this sinusoidal curve and a horizontal line at $2 \chi/h_{\rm k}$, and
it increases with increase of $2 \chi/h_{\rm k}$ as shown in
Fig. \ref{fig:FixedPoint}(b). In Fig. \ref{fig:FixedPoint}(b), the
curves represent the analytical results obtained by 
Eq. \eqref{eq:Sin2phiSinphi} and the symbols represent the numerical
results obtained by directly solving the Eqs. \eqref{eq:LLGtheta} and
\eqref{eq:LLGphi} with $\alpha=0.02$,
$h_{\rm k}=0.01$, and 
$k_{\rm u}^{\rm eff} = -0.4$.
The analytical and simulation
results agree very well with each other.  We also performed numerical
simulations for wide range of $\alpha$ and confirmed that the numerical
results of $\phi_{0}$ are independent of $\alpha$ as predicted by the
analytical results. In the numerical simulations, the current density 
was gradually increased from zero. At each current
density, the simulation was run long enough for the polar and azimuthal
angles to be converged to $\theta_{0}$ and $\phi_{0}$.

Numerical simulations showed that there exists a critical
current density, $J_{\rm c 1}$, above which the OP precession is induced.
For $J>0$, $J_{\rm c 1}$ is obtained by calculating the
maximum value ($\Lambda$) of the left hand side (LHS) of
Eq. \eqref{eq:Sin2phiSinphi}. If $2 \chi/h_{\rm k}$ is larger than $\Lambda$, 
there is no fixed point and the limit cycle corresponding to the OP
precession is induced.  Hereafter we consider the case of $J>0$, 
however, the critical current density for $J<0$ can be obtained in the
similar way by calculating the minimum value.

\begin{figure}[t]
%\begin{figure}[P]
\includegraphics[width=0.95\columnwidth]{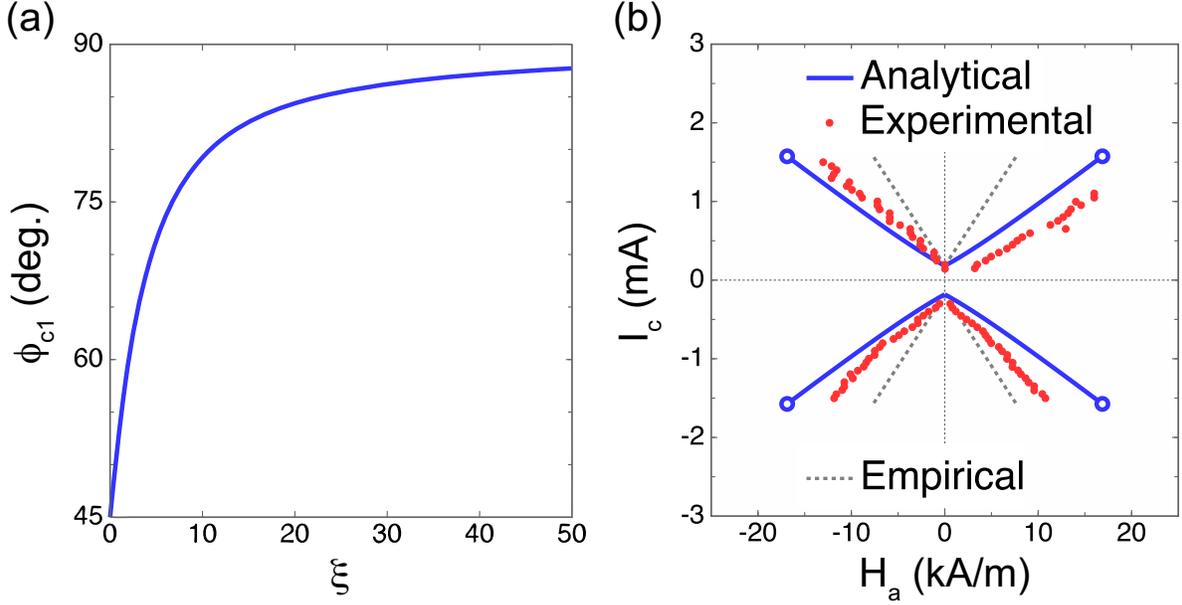}% Here is how to import EPS art
\caption{\label{fig:Jc}
  (a) Analytically-calculated $\xi$ dependence of critical $\phi$ ($\phi_{\rm c1}$). 
  $\xi$ is ratio between 
  $H_{\rm a}$ and IP shape anisotropy field ($H_{\rm a}$), being
  $\xi=2H_{\rm a}/H_{\rm a}$.
  (b) $H_{\rm a}$ dependence of 
  critical current ($I_{\rm c}$) for OP precession.
  Solid blue curves represent plots of analytical expression (Eq. (\ref{eq:Jc1})).
  $H_{\rm k}$ of 4 kA/m is assumed. 
  Open blue circles represent critical current 
  above which the OP precession can not be maintained.
  Red dots represent past experimental results 
  (redrawn from Ref.    
   \citenum{houssameddine_spin-torque_2007}).
  Dotted gray lines represent the empirically approximated value proposed in 
  Ref. \citenum{houssameddine_spin-torque_2007}. 
}
\end{figure}

%========================================
%  Analytical expression of Jc
%========================================
At the maximum, the derivative of 
the LHS of Eq. \eqref{eq:Sin2phiSinphi} with respect to $\phi_{0}$
is zero, that is,
\begin{align}
\label{eq:Eqforphic}
 2\cos 2\phi_{0}  + \xi \cos \phi_{0}  = 0.
\end{align}
Expressing cosine functions by $\tan\phi_{0}$, 
one can easily obtain the 
solution of Eq. (\ref{eq:Eqforphic}) as
\begin{align}
\label{eq:phic}
\phi_{\rm c 1} = \arctan \left[\frac{1}{2\sqrt{2}} \sqrt{ \xi^{2} + 8 +\xi \sqrt{\xi^{2} +32} } \right],
\end{align}
where the subscript ``c1'' stands for the critical value corresponding to
$J_{\rm c 1}$.  Fig. \ref{fig:Jc}(a) shows $\xi$ dependence of $\phi_{\rm c1}$
given by Eq. \eqref{eq:phic}. $\phi_{\rm c1}= \pi/4$ for $\xi=0$, i.e.,
$h_{\rm a}=0$.
It monotonically increases with increase of $\xi$ and reaches $\pi/2$ in
the limit of $\xi \rightarrow \infty$, i.e., $h_{\rm k} \to 0$.

The maximum value, $\Lambda$, can be obtained by substituting
$\phi=\phi_{\rm c1}$ into the LHS of Eq. \eqref{eq:Sin2phiSinphi}
as
\begin{align}
\label{eq:LocalMax}
\Lambda = \frac{ \sqrt{ X + 8} \left( \xi \sqrt{ X + 16 } + 4 \sqrt{2} \right) }{ X + 16 },
\end{align}
where $X= \xi (\xi  + \sqrt{ \xi^{2} + 32 }) $.
Equating this maximum value with $2\chi/h_{\rm k}$ and
using Eq. \eqref{eq:chi}, the critical current density is obtained as
\begin{align}
\label{eq:Jc1}
  J_{\rm c 1}
  =
  \frac{e \mu_{0} M_{\rm s} d H_{\rm k}}{\hbar P}
 \frac{ \sqrt{ X + 8} \left( \xi \sqrt{ X + 16 } + 4 \sqrt{2} \right) }{ X + 16 }.
\end{align}
This is the main result of this letter.
It should be noted $J_{\rm c 1}$ is also independent of $\alpha$. 
In the absence of the applied IP magnetic field,
i.e., $H_{\rm a} = 0$, Eq. \eqref{eq:Jc1} becomes
\begin{align}
\label{eq:chic2}
  J_{\rm c 1}\bigl|_{H_{\rm a}=0}
  =
  \frac{e \mu_{0} M_{\rm s} d H_{\rm k}}{\hbar P}.
\end{align}
In the limit of $H_{\rm k}\to 0$,  it reduces to
\begin{align}
\label{eq:chic3}
  \lim_{H_{\rm k}\to 0}
  J_{\rm c 1}
  =
  \frac{2e \mu_{0} M_{\rm s} d H_{\rm a}}{\hbar P}.
\end{align}
For small magnetic field such that $H_{\rm a} \ll  H_{\rm k}$, i.e.,
$\xi \ll 1$, it can be approximated as
\begin{align} 
  \label{eq:chi_approx}
  J_{\rm c 1}
  \simeq
  \frac{e \mu_{0} M_{\rm s} d}{\hbar P}
  \left(H_{\rm k} + \sqrt{2} H_{\rm a}\right), 
\end{align}
by noting that the Taylor expansion of $\Lambda$ around $\xi=0$ is given
by $\Lambda = 1+ \xi/\sqrt{2} + \xi^{2}/16 + O(\xi^{3}) $.

%==============================
% Fig. 3
%==============================

%========================================
%  The other fixed point
%========================================
Once the current density, $J$,  exceeds $J_{\rm c1}$, the OP precession is
excited and further increase of $J$ moves the trajectory towards the south
pole ($\theta=\pi$). Around $\theta=0$ and $\pi$, there exist the fixed points
other than $\theta_{0}=\pi/2$, which are determined by
\begin{align}
&
 2 \sin \theta_{0}  (\frac{h_{\rm k}}{2} \cos^{2} \phi_{0} - k_{\rm u1}^{\rm
   eff}) +  h_{\rm a} \cos \phi_{0} =0,
\\
&
\frac{h_{\rm k}}{2} \sin \theta_{0} \sin 2\phi_{0} +  h_{\rm a} \sin \phi_{0}
=\chi\sin\theta_{0}.
\end{align}
After some algebra, the fixed point is obtained as
\begin{align}
  \label{eq:fp2_theta}
  &\theta_{0}
  =
  \arcsin\left(
  \frac{
    h_{\rm a}
    \sqrt{
      \left(
        k_{\rm u1}^{\rm eff}
      \right)^{2}
      +\chi^{2}
    }
  }
  {
  2
  \left[
  \left(
  k_{\rm u1}^{\rm eff}
  \right)^{2}
  +\chi^{2}
  \right]
  -
  h_{\rm k}k_{\rm u1}^{\rm eff}
  }
  \right),
  \\
  \label{eq:fp2_phi}
  &\phi_{0}
  =
  -
  \arctan
  \frac{\chi}{k_{\rm u1}^{\rm eff}},
\end{align}  
where $\pi/2 < \left|\phi_{0}\right|\le\pi$. In the absence of the
applied IP magnetic field, i.e., $h_{\rm a}= 0$, the polar angle of the fixed point
is $\theta_{0} = 0$ or $\pi$. It is difficult to obtain the exact
analytical expression for the critical current density, $J_{\rm
c 2}$, above which the OP precession can not be maintained, and $\bm{m}$
stays at the fixed point given by Eqs. \eqref{eq:fp2_theta} and
\eqref{eq:fp2_phi}.
Since the average polar angle of the trajectory of the OP precession is
determined by the competition between the damping torque and spin torque,
this critical current density should depend on $\alpha$.  The
approximate expression was obtained by Ebels 
$et$ $al$.
\cite{ebels_macrospin_2008} as

\begin{align} 
  \label{eq:jc2_approx}
  J_{\rm c 2}
  \simeq
  -\frac{4\alpha e d K_{\rm u1}^{\rm eff}}{\hbar P},
\end{align}
which agrees well with the macrospin simulation results.

%========================================
%  Comparison: Analytical calculation vs Experiment
%========================================
Let us compare our results with the experimental results 
reported by D. Houssameddine $et$ $al$.
\cite{houssameddine_spin-torque_2007}
Figure \ref{fig:Jc}(b) shows the applied IP magnetic field, $H_{\rm a}$, dependence of 
critical current ($I_{\rm c}$) for the OP precession.
The analytical results of Eq. \eqref{eq:Jc1} are plotted 
by the solid (blue) line and the experimental results are plotted by the (red) dots.
The critical current corresponding to $J_{\rm c 2}$
%$\chi \sim 2 \alpha k_{\rm u1}^{\rm eff}$
%\cite{ebels_macrospin_2008} 
are also shown by open (blue) circles.
In the analytical calculation, the following parameters indicated in
Ref. \citenum{houssameddine_spin-torque_2007} are assumed:
$\alpha=0.02$, $M_{\rm s}=866$ kA/m, 
the junction area is $30 \times 35 \times \pi$ nm$^{2}$,  $d=3.5$ nm,
$P=0.3$, $H_{\rm k}=4$ kA/m.
The dotted (gray) lines represent the approximated values
proposed in 
Ref. \citenum{houssameddine_spin-torque_2007}, 
$I_{\rm c} \propto H_{\rm k} +  2 H_{\rm a}$.
One can clearly see that the analytical results of
Eq. \eqref{eq:Jc1} reproduces the experimental results
very well. The agreement is much better than the approximated values of
Ref. \citenum{houssameddine_spin-torque_2007}.
As shown in Eq. \eqref{eq:chi_approx}, the critical current for small
magnetic field can be approximated as  $I_{\rm c} \propto H_{\rm
k} + \sqrt{2}H_{\rm a}$ rather than $I_{\rm c} \propto H_{\rm k} + 2 H_{\rm a}$.

%========================================
%  Conclusion
%========================================
In summary,  
we theoretically studied 
spin-torque induced magnetization dynamics
in an STO with an IP magnetized free layer 
and an OP magnetized polarizer.
We obtained the rigorous analytical expressions of 
$J_{\rm c1}$ for the
OP precession in the presence of
IP shape-anisotropy field ($H_{\rm k}$) and applied IP
magnetic field ($H_{\rm a}$).
The expression reproduces the experimental results very
well and revealed that the critical current is proportional to 
$H_{\rm k} + \sqrt{2}  H_{\rm a}$ for $H_{\rm a} \ll H_{\rm k}$.

\begin{acknowledgments}
This work was supported by JSPS KAKENHI Grant Number 16K17509.
\end{acknowledgments}

%\nocite{*}
%\bibliographystyle{apsrev4-1}
%\bibliography{ref}% Produces the bibliography via BibTeX.

\providecommand{\noopsort}[1]{}\providecommand{\singleletter}[1]{#1}%

\end{document}